\documentclass{aa}
\usepackage{txfonts}
\usepackage{graphicx}
\usepackage{natbib}
\bibpunct{(}{)}{;}{a}{}{,}

\begin{document}

\title{XMM-{\it Newton} observations of the $\sigma$~Ori cluster. \\
I. The complex RGS spectrum of the hot star $\sigma$~Ori~AB} 
\author{J. Sanz-Forcada\thanks{\em{Present address:} 
	Astrophysics Division -- Research and Science Support Department
  	of ESA, ESTEC, Postbus 299, NL-2200 AG Noordwijk, The Netherlands} 
	\and E. Franciosini \and R. Pallavicini}
\institute{INAF - Osservatorio Astronomico di Palermo,
Piazza del Parlamento 1, I-90134 Palermo, Italy}

\offprints{J. Sanz-Forcada, \email{jsanz@rssd.esa.int}}

\date{Received 29 January 2004 / Accepted 30 March 2004}
\titlerunning{XMM-{\it Newton} observations of the $\sigma$~Ori cluster. I.}

\abstract{
We present XMM-{\it Newton} observations of the young ($\sim 2-5$ Myr)
cluster around the hot (O9.5V) star $\sigma$~Orionis~AB, aimed at obtaining
a high resolution RGS spectrum of the hot star as well as EPIC imaging data
for the whole field. We show that the RGS spectrum of $\sigma$~Ori~AB may be
contaminated by weaker nearby sources which required the development of a
suitable procedure to extract a clean RGS spectrum and to determine the
thermal structure and wind properties of the hot star. We also report on the
detection of a flare from the B2Vp star $\sigma$~Ori~E and we discuss whether
the flare originated from the hot star itself or rather from an unseen
late-type companion. Other results of this observation include: the
detection of 174 X-ray sources in the field of $\sigma$~Ori of which 75
identified as cluster members, including very low-mass stars down to the
substellar limit; the discovery of rotational modulation in a late-type star
near $\sigma$~Ori~AB; no detectable line broadenings and shifts ($\la
800$~km~s$^{-1}$) in the spectrum of $\sigma$~Ori~AB together with a
remarkable low value of the \ion{O}{vii} forbidden to intercombination line
ratio and unusually high coronal abundances of CNO elements.

}
\maketitle


\section{Introduction}

The $\sigma$~Ori cluster, discovered by {\it ROSAT} \citep{wolk96, walter97}
around the O9.5V star $\sigma$~Ori~AB, belongs to the OB1b association and
is located at a distance of $352_{-85}^{+166}$ pc \citep[from {\it
Hipparcos},][]{hippa}. In addition to several hot stars, it is known to
contain $\sim 100$ likely pre-main sequence late-type stars within
$30^\prime$ of $\sigma$~Ori, as well as some brown dwarfs and planetary-mass
objects \citep{bejar99,zapat00,bejar01}. The estimated age of the cluster is
$2-5$ Myr.

We have obtained an XMM-{\it Newton} observation of the $\sigma$~Ori
cluster, centered on the hot star $\sigma$~Ori~AB, with the purpose of
obtaining: i) a high-resolution RGS spectrum of the central source; ii)
imaging data as well as low-resolution spectra over the whole field,
including both a few early-type stars and a large number of late-type stars
down to the substellar limit. Given the high sensitivity of XMM-{\it
Newton}, and the good combination of low- and high-resolution spectroscopic
instruments on board, these observations were expected to shed light on the
coronal and/or wind properties of stars in a very young cluster.

X-ray emission from O and B stars is usually explained with the presence of
winds. X-ray observations of the hot stars $\zeta$~Pup (O4If) and
$\zeta$~Ori (O9Ib) have shown the presence of such winds, with velocity
widths of the order of $600-1500$~km~s$^{-1}$ and blueshifted centroids
\citep{wald01,cas01,kahn}. However, some conflicting results have also been
found, like high-densities close to the stellar surface in $\zeta$~Ori
\citep{wald01}, where the velocity is too small to produce the shocks
required for the X-ray emission, or a temperature structure in the Orion
Trapezium hot stars that is similar to that of cool active stars, where the
emission originates from magnetically confined coronal structures
\citep{sch03}. This has raised the question of whether coronal loops might
be present in some hot stars partially contributing to their X-ray emission.
High-resolution spectroscopic observations of the hot star $\sigma$~Ori with
XMM-{\it Newton} can clarify some of these issues.

As it will be described below, there are several X-ray sources, both hot and
cool stars, in our XMM-{\it Newton} field close enough to the central source
to potentially contaminate its high-resolution RGS spectrum. Although their
X-ray intensity lies well below the level of $\sigma$~Ori~AB, they produce
lines that can contribute significantly at certain wavelengths. These lines
are shifted in wavelength with respect to those emitted by $\sigma$~Ori~AB,
because of the different locations of these sources within the RGS field of
view (FOV). These spurious lines must be identified in order to correctly
analyze the spectrum of the central source. To this aim, the capability of
XMM-{\it Newton} of obtaining simultaneous high-resolution spectra of the
central source with RGS and low-resolution spectra of the other sources in
the field with EPIC is of great advantage. By using the information derived
from the EPIC spectra, it is possible in fact to model the expected
contributions of nearby sources to the RGS spectra. This, together with the
wavelengths shifts in the RGS spectra caused by the different offsets of the
sources in the RGS FOV, allows an accurate correction of the $\sigma$~Ori~AB
spectrum for the contribution of nearby sources.

\begin{figure}
\resizebox{\hsize}{!}{\includegraphics[clip]{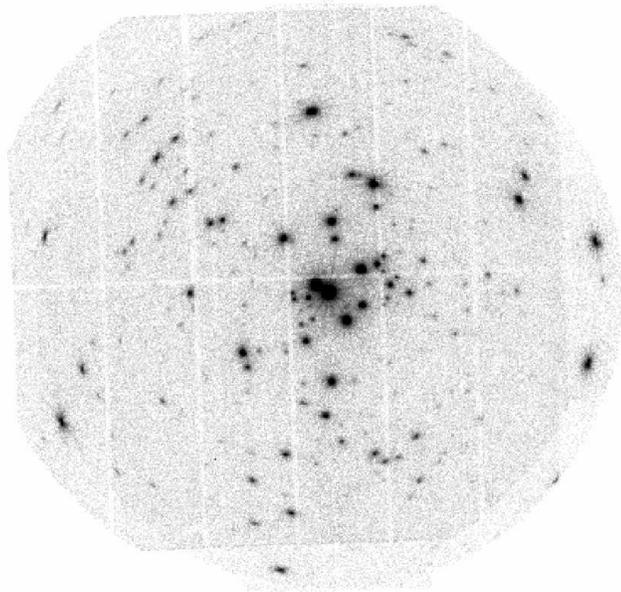}}
\caption{Composite EPIC (MOS1+MOS2+PN) image of the $\sigma$~Ori cluster}
\label{fig:sori_ima}
\end{figure}

This paper is organized as follows. In Sect.~\ref{sec:observ} we present the
analysis of the XMM-{\it Newton} observation, discussing first the imaging
data obtained with the EPIC PN and MOS detectors, and then the
high-resolution spectroscopic data obtained with the RGS. Since the observed
RGS spectra of the central source might in principle be contaminated by up
to three nearby sources in addition to $\sigma$~Ori~AB itself, we present in
this section light curves and low-resolution EPIC spectra of these four
sources, showing that one of them (the hot star $\sigma$~Ori~E) is
undergoing a flaring episode, another one (a K star) is showing evidence of
rotational modulation, while $\sigma$~Ori~AB and the fourth source (another
K star) are either quiescent or of low variability. With regard to the EPIC
spectra also presented in this section, $\sigma$~Ori~AB appears much softer
than the other three sources, consistently with typical X-ray spectra of hot
stars. In Sect.~\ref{sec:results}, we present the results of our analysis,
first for $\sigma$~Ori~AB and then for the flare on $\sigma$~Ori~E. In
particular, we derive, for the former star, the ``cleaned'' differential
emission measure distribution, and we put constraints on wind velocities and
shifts, chemical abundances and densities. For $\sigma$~Ori~E we discuss the
flare properties and we present evidence for circumstellar absorption. In
Sect.~\ref{sec:disc}, we discuss the implication of our results, both for
current models of shocked winds in early-type stars (as in $\sigma$~Ori~AB),
and for the possible occurence of flares in hot stars (as opposite to flares
originating from unseen late-type companions). Finally in
Sect.~\ref{sec:conclusions} we summarize our conclusions.


\section{Observations}\label{sec:observ}

XMM-{\it Newton} observations of the $\sigma$~Ori cluster, centered on the
hot star $\sigma$~Ori~AB, were carried out as part of the Guaranteed Time of
one of us (R.P.) from 21:47 UT on March 23, 2002 to 9:58 UT on March 24,
2002 (obs. ID 0101440301), for a total duration of 43 ks. We used both the
EPIC \citep[European Photon Imaging Camera,][]{tur01,str01} instrument with
PN and MOS detectors (sensitive to the spectral range 0.15--15 keV and
0.2--10~keV, respectively) and the RGS \citep[Reflection Grating
Spectrometer,][]{denher01} instrument (sensitive to the range
$\lambda\lambda\sim 6-38$~\AA). This allowed us to obtain simultaneously
low-resolution CCD spectra ($\Delta E\sim 70$~eV at $E\sim 1$~keV) of the
brightest sources in the field, and high-resolution grating spectra
($\lambda/\Delta\lambda\sim 100-500$) of $\sigma$~Ori~AB. The EPIC cameras
were operated in Full Frame mode using the thick filter. Data analysis was
carried out using the standard tasks in SAS v.5.4.1

\subsection{EPIC observations of the $\sigma$~Ori field}
\label{sec:epicfield}

EPIC calibrated and cleaned event files were derived from the raw data using
the standard pipeline tasks {\sc emchain} and {\sc epchain} and then
applying the appropriate filters to eliminate noise and bad events. The
event files have also been time filtered in order to exclude a few short
periods of high background due to proton flares; the final effective
exposure time is 41 ks for each MOS and 36 ks for the PN. The combined EPIC
(MOS1+MOS2+PN) image in the $0.3-7.8$ keV energy band is shown in
Fig.~\ref{fig:sori_ima}.

Source detection was performed both on the individual datasets and on the
merged MOS1+MOS2+PN dataset using the Wavelet Detection algorithm developed
at the Osservatorio Astronomico di Palermo \citep[Damiani et al., in
preparation]{damiani97}. For the detection on the summed dataset, a combined
exposure map has been computed by summing the individual exposure maps with
an appropriate scaling factor for PN, derived from the median ratio of PN to
MOS count rates, in order to take into account the different sensitivities
of MOS and PN. For our observation the median PN/MOS ratio is $\sim 3.2$,
resulting in a MOS equivalent exposure time of $\sim 200$ ks for the merged
dataset. We detected a total of 174 sources above a significance threshold
of $5\sigma$. We have identified 75 sources with at least one possible
cluster member or candidate within $10\arcsec$ of the X-ray position. Of the
detected members, 5 are early-type stars, including $\sigma$~Ori~AB and E,
while 7 sources have been identified with very low-mass stars of spectral
type later than $\sim\,$M5. Among the latter ones is SOri~68, a planetary-mass
object of spectral type L5.0 \citep{bejar01} and the candidate brown dwarf
SOri~25, which has a spectral type M6.5 and an estimated mass of $0.05-0.13 \,
M_\odot$ \citep{bejar99}.

A more detailed analysis of the full EPIC field will be presented in a
companion paper (Franciosini et al., in preparation). In the following we
will concentrate only on the brightest central sources that might contribute
significantly to the RGS spectra.

\begin{figure}
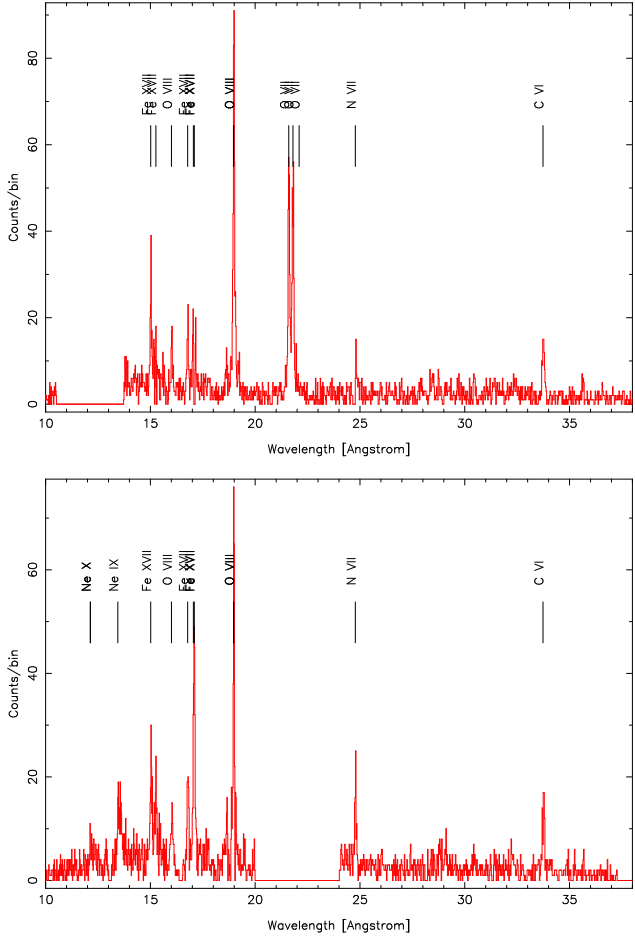

\resizebox{\hsize}{!}{\includegraphics[angle=270]{pallavicini_r_fig3a.ps}}
\resizebox{\hsize}{!}{\includegraphics[angle=270]{pallavicini_r_fig3b.ps}}
\caption{RGS 1 \& 2 spectra of the central source in the $\sigma$~Ori field. 
The most intense lines identified are marked. The observed spectrum is
produced by the hot star $\sigma$~Ori~AB plus some contribution from nearby
sources (mainly the flaring source $\sigma$~Ori~E) falling in the RGS FOV.
\label{fig:rgsspec}}
\end{figure}

\begin{figure}
\resizebox{\hsize}{!}{\includegraphics[clip]{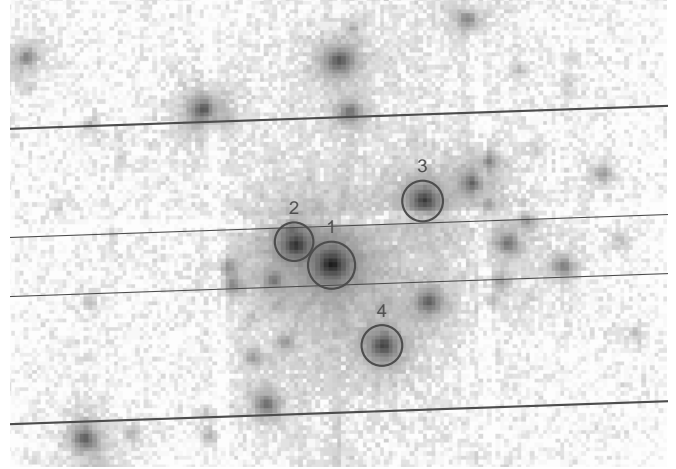}}
\caption{Central part ($11\arcmin \times 8\arcmin$) of the composite EPIC
image of the $\sigma$~Ori field. Circles mark the sources falling in the RGS
FOV with an EPIC flux $>20\%$ of that of $\sigma$~Ori~AB (all other sources
in the FOV contribute less than 10\% of the $\sigma$~Ori~AB flux); these
sources are listed in Table~\ref{tab:sources} with the corresponding
identification numbers. The RGS dispersion direction runs parallel to the PN
chip separation (which in this case is almost coincident with the right
ascension axis), with wavelength increasing to the left. Thick lines mark
the limits of the RGS FOV (5$\arcmin$ wide), and thin lines the limits of
the extraction region for the RGS spectra. 
\label{fig:sigorifov}}
\end{figure}

\begin{figure}
\resizebox{\hsize}{!}{\includegraphics[clip]{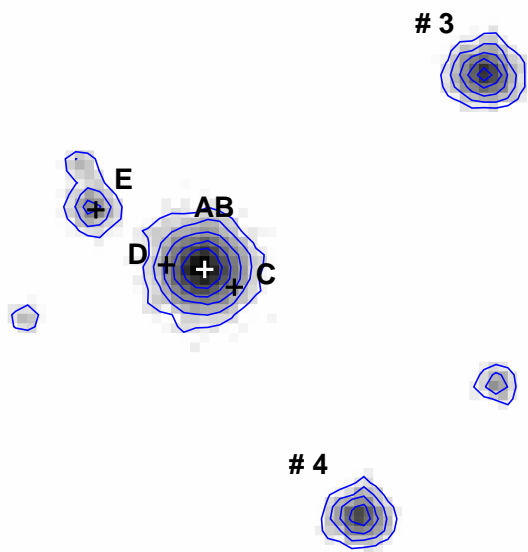}}
\caption{Close-up view of the central region of the $\sigma$~Ori field
during the time interval before the $\sigma$~Ori~E flare,
showing the position of the components of the $\sigma$~Ori system.
Contours are at 2.5, 5, 10, 20, 40 cts/arcsec$^2$. Note the weak source
to the north of $\sigma$~Ori~E}
\label{fig:sori_comp}
\end{figure}

\begin{table*}
\caption{Sources falling in the RGS field of view with an EPIC flux more
than 20\% that of $\sigma$~Ori~AB. Count rates are MOS equivalent count
rates. Identified stars are all members or candidate members of the
$\sigma$~Ori cluster 
\label{tab:sources}}
\begin{tabular}{cccrlcrrl}
\hline\hline
$N_x$& RA$_x$& DEC$_x$& count rate& Opt. ID& offset& $V$& $B-V$& Sp. type \\
& \multicolumn{2}{c}{(2000)}& (cts/ks)& & ($\arcsec$)& & & \\ 
\hline
1& $5\>38\>44.86$& $-2\>36\>00.2$& $435.9 \pm 2.4$& 
  $\sigma$~Ori~AB& 1.17& 3.78& $-0.24$& O9.5V\\
2& $5\>38\>47.26$& $-2\>35\>39.9$& $197.6 \pm 1.7$& 
  $\sigma$~Ori~E & 1.08& 6.70& $-0.18$& B2Vp \\
3& $5\>38\>38.54$& $-2\>34\>55.5$& $137.6 \pm 1.4$& 
  GSC 4771-1147  & 0.91& 12.24&  $1.18$& K5  \\
4& $5\>38\>41.34$& $-2\>37\>22.3$& $106.2 \pm 1.2$& 
  r053841-0237   & 0.82&     &      & K3 \\
\hline
\end{tabular}
\end{table*}

\subsection{RGS observations of $\sigma$~Ori~AB}
\label{sec:rgsobs}

RGS spectra were extracted for the source located at the center of the EPIC
field, which coincides with the position of the hot star $\sigma$~Ori~AB.
The extracted RGS1 and RGS2 spectra have an effective exposure time of 42
ks; they are shown in Fig.~\ref{fig:rgsspec}, where we also indicate the
identification of the most prominent lines.

The RGS instrument has a rectangular field of view (FOV), with the longer
axis along the dispersion direction and a width of $5\arcmin$ in the cross
dispersion direction. The dispersion direction is parallel to the separation
of the upper and lower chips in the PN detector, with wavelength increasing
towards the left; in our case it is almost aligned along the right ascension
coordinate, as indicated in Fig.~\ref{fig:sigorifov}. The figure shows that
there are other bright X-ray sources falling in the RGS FOV that might in
principle contribute to the observed spectrum of the central source. From
their EPIC fluxes, we see that there are 3 sources, listed in
Table~\ref{tab:sources}, with fluxes ranging from 24\% to 45\% of the
$\sigma$~Ori~AB flux, while all other sources in the FOV have fluxes less
than 10\% of that of $\sigma$~Ori~AB. However, from the size of the
extraction region used to derive the RGS spectrum (indicated by thin lines
in Fig.~\ref{fig:sigorifov}), we see that only $\sigma$~Ori~E should
contaminate the spectrum of $\sigma$~Ori~AB, while sources \#3 and
especially \#4 should not give any significant contribution. 

The presence of emission from a source with a certain offset $\theta$ in
right ascension with respect to the center of the field produces a shift in
the wavelength scale of RGS according to $\Delta \lambda=0.124 \times
\theta/m$, where $\lambda$ is measured in \AA, $\theta$ in arcmin, and $m$
is the RGS spectral order. Hence, it is important to isolate the
contribution of each source to the spectrum, especially considering the fact
that intense lines of a secondary source may fall at wavelengths where weak
lines of the main source are. The wavelength shifts, due to the different
positions of the sources in the RGS FOV, and the knowledge of their
low-resolution spectra from EPIC, make it possible to identify the
contaminating lines and to estimate their contribution to the RGS spectrum of
$\sigma$~Ori~AB, as we will show in Sect.~\ref{sec:rgsAB}. 

The central source $\sigma$~Ori is a remarkable quintuple system. Components
A and B are separed by only 0.25 arcsec, and are therefore unresolved. The
other 3 components ($\sigma$~Ori~C, D and E) are at distances of 11.2, 12.9
and 42 arcsec, respectively: except for $\sigma$~Ori~E, they cannot be
easily resolved from $\sigma$~Ori~AB with XMM-{\it Newton}. In
Fig.~\ref{fig:sori_comp} we show a close-up view of the center of the field
in the time interval preceding the $\sigma$~Ori~E flare (see
Sect.~\ref{sec:soriE_flare}), with the positions of the 5 components marked.
The figure clearly shows that the contribution of $\sigma$~Ori~C and D to
the X-ray emission of the central source is negligible (note, for
comparison, that the quiescent flux of $\sigma$~Ori~E is only $\sim 9$\% of
the flux of $\sigma$~Ori~AB). This conclusion is supported by a higher
resolution {\it Chandra} observation of the same field by \citet{wolk04}. At
the resolution of {\it Chandra}, components C and D are easily resolved from
components A and B: however, the {\it Chandra} image shows only a very weak
X-ray source ($<1$\% of the $\sigma$~Ori~AB flux) at the position of
$\sigma$~Ori~D (a B2V star), and no emission at the position of
$\sigma$~Ori~C (a A2V star). The {\it Chandra} observation reveals another
X-ray source $\sim 2\arcsec$ to the NW of $\sigma$~Ori~AB, which appears to
be associated with the IR object $\sigma$~Ori~IRS1 \citep{loo03}, attributed
to emission from a protoplanetary disk. This source cannot be resolved at
the lower resolution of XMM-{\it Newton} and thus could potentially
contaminate both the EPIC and RGS spectra of $\sigma$~Ori~AB. From the {\it
Chandra} image, we estimate however that its contribution is $\sim 3$\% of
the flux of $\sigma$~Ori~AB and is thus negligible.

\begin{figure*}
\resizebox{\hsize}{!}{\includegraphics[clip]{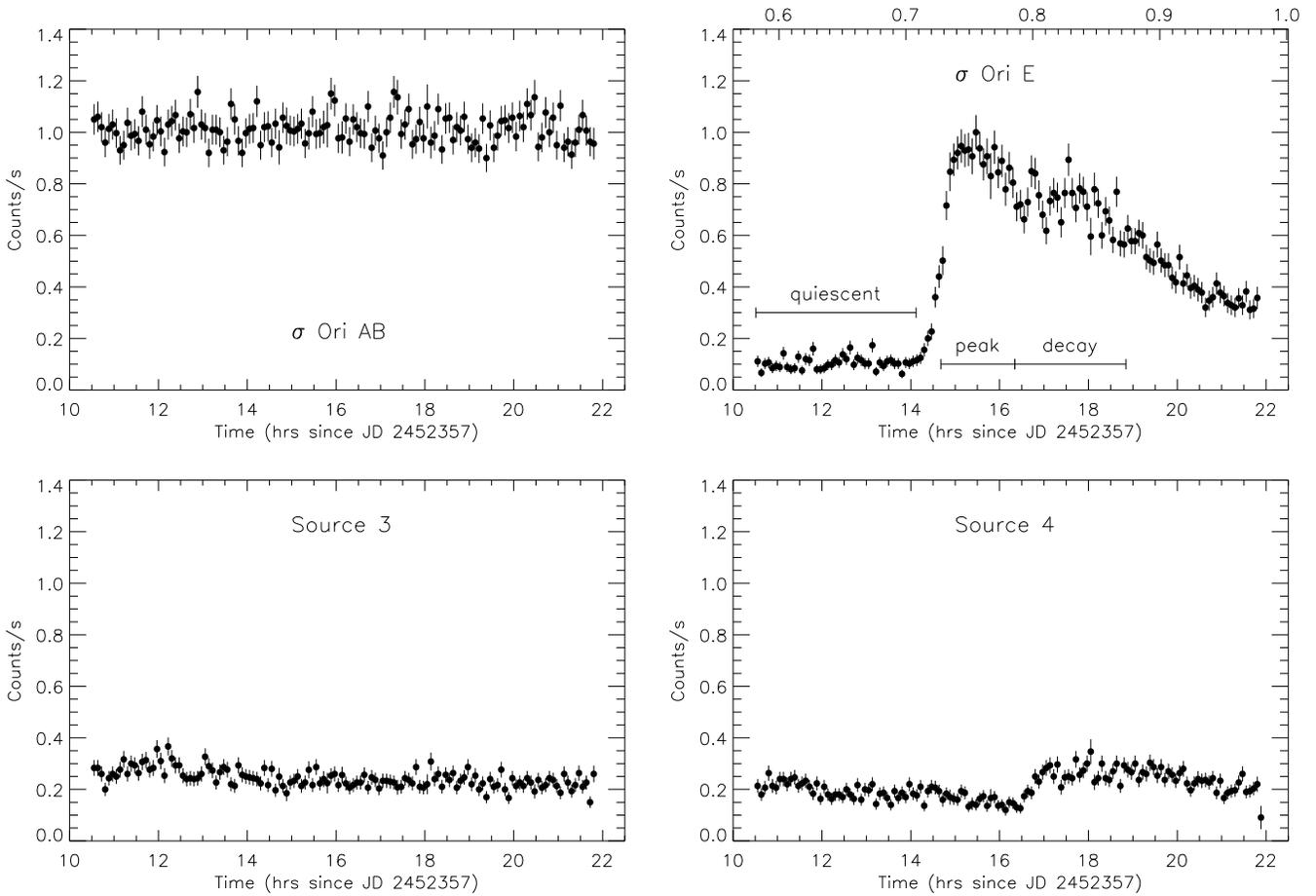}}
\caption{EPIC PN light curves of the 4 sources listed in
Table~\ref{tab:sources}, binned over 300 sec. In the case of $\sigma$~Ori~E,
the upper axis indicates the photometric phase according to \citet{rei00},
with phase 0 indicating the maximum absorption in the U-band (the secondary
minimum ends at $\Phi=0.5$). We also show the time intervals used for the
time-resolved spectral analysis of $\sigma$~Ori~E 
\label{fig:lcurves}}
\end{figure*}

\subsection{EPIC light curves and spectra of the central sources}
\label{sec:centr_src}

EPIC PN light curves for the four main sources in the central region of the
field were extracted using a circular region of radius $24\arcsec$, except
for $\sigma$~Ori~E where we have used a radius of $18\arcsec$ to avoid
contamination from $\sigma$~Ori~AB; in the latter case we also excluded a
small region around a faint nearby source detected to the north of
$\sigma$~Ori~E ($\alpha = 5\;38\;47.56$, $\delta = -2\;35\;24.7$, see
Fig~\ref{fig:sori_comp}). The four light curves, binned over 300~s, are
shown in Fig.~\ref{fig:lcurves}.


As shown by Fig.~\ref{fig:lcurves}, $\sigma$~Ori~AB is clearly the dominant
source and its emission is steady (the $1\sigma$ standard deviation around
the mean count rate is $\la$\,1\%). The other hot star ($\sigma$~Ori~E) shows
instead the occurrence of a flare, with a factor of 10 increase in the count
rate. The occurrence of a flare in a hot star is at variance with current
models of X-ray emission in early-type stars: this, as well as the
properties of the flare, will be discussed in Sect.~\ref{sec:soriE_flare}.
The other two sources (\#3 and \#4) are identified with K-type candidate
members of the cluster and their X-ray emission, which shows a quite high
level of variability, is likely due to magnetically-confined coronal
structures. Particularly interesting is source \#4 which shows evidence of
rotational modulation, with a period of $\sim 8.5$ hours and an amplitude of
$\sim 25$\% \citep[see their Fig.~3]{pallavic04}, that can be attributed to
the inhomogeneous distribution of active regions over the surface of the star.

\begin{figure*}
\resizebox{\hsize}{!}{\includegraphics[angle=90]{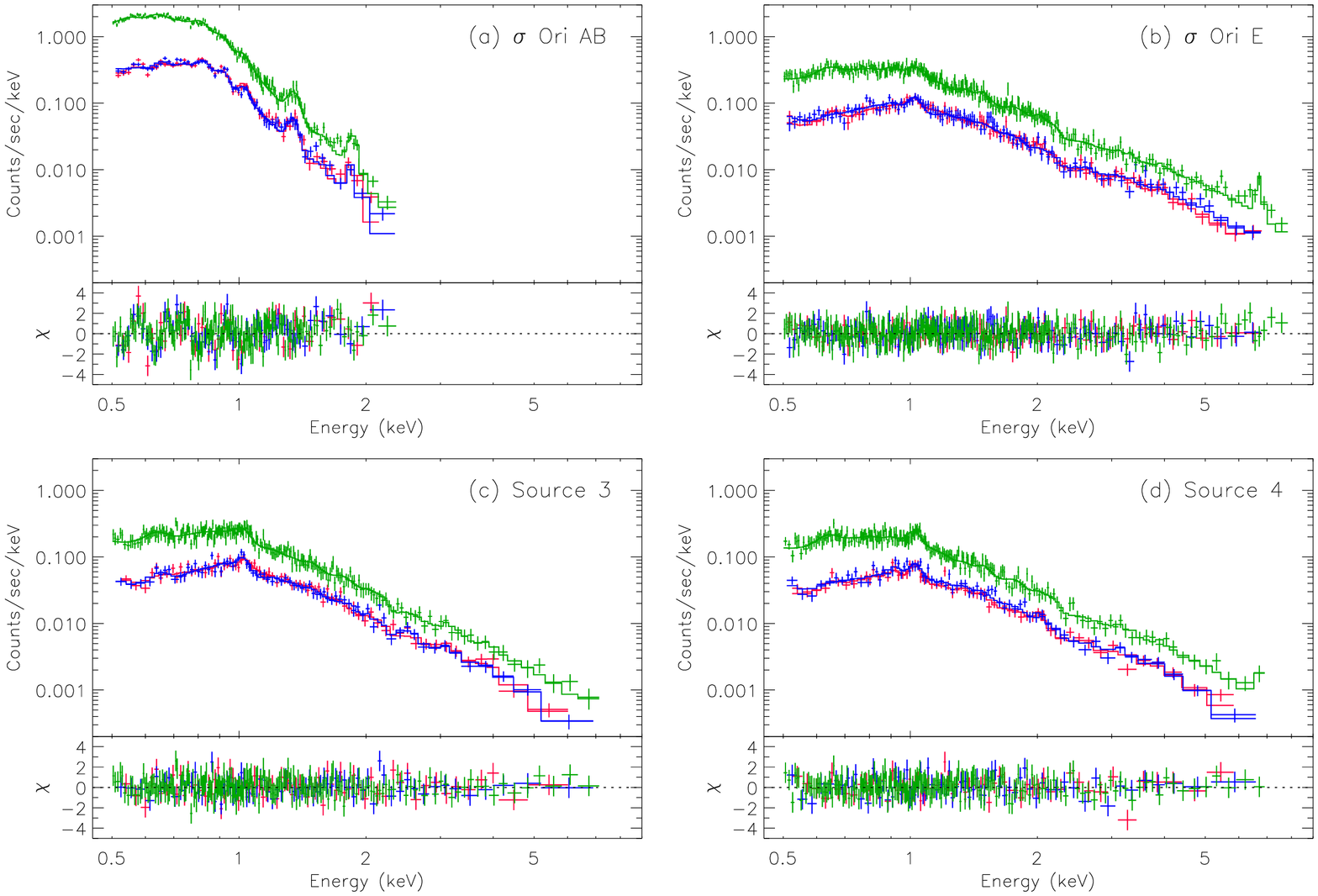}} 
\caption{EPIC MOS1 ({\it blue}), MOS2 ({\it red}) and PN ({\it green})
spectra of the 4 sources listed in Table~\ref{tab:sources}. The best-fit
models and the residuals are also shown. The spectrum of $\sigma$~Ori~AB is
dominated by background above 2.5~keV, while the spectrum of $\sigma$~Ori~E
refers to the total (quiescent $+$ flare) observation.
\label{fig:epicspec}}
\end{figure*}

\begin{table*}
\caption{Best-fit parameters for the EPIC spectra of the four sources in
Table~\ref{tab:sources}. Note that the fit to $\sigma$~Ori E refers to the
global (quiescent+flare) spectrum. Errors are 90\% confidence ranges for one
interesting parameter}
\begin{tabular}{llllclll}
\hline\hline\noalign{\smallskip}
  &  MOS1 & MOS2& PN & \qquad\qquad & MOS1 & MOS2& PN\\
\hline \noalign{\smallskip}
 &\multicolumn{3}{c}{\bf $\sigma$~Ori~AB} & &
\multicolumn{3}{c}{\bf $\sigma$~Ori~E} \\[6pt]
$T_1$ (keV)& 
 $0.13_{-0.05}^{+0.07}$& $0.13_{-0.05}^{+0.09}$& $0.17_{-0.07}^{+0.31}$ & &
 $0.75_{-0.20}^{+0.11}$& $0.55_{-0.13}^{+0.23}$& $0.41_{-0.04}^{+0.05}$\\[3pt]
$T_2$ (keV)& 
 $0.35_{-0.02}^{+0.04}$& $0.32_{-0.02}^{+0.02}$& $0.33_{-0.02}^{+0.10}$ & &
 $2.96_{-0.31}^{+0.36}$& $3.83_{-0.84}^{+2.54}$& $3.24_{-0.20}^{+0.23}$\\[3pt]
$EM_1$ ($10^{53}$ cm$^{-3}$)& 
 $15.7_{-7.3}^{+5.9}$  & $21.5_{-8.4}^{+7.1}$  & $19.4_{-2.2}^{+27.1}$ & &
 $2.52_{-1.02}^{+2.15}$& $11.2_{-8.2}^{+11.8}$ & $2.76_{-0.59}^{+0.81}$\\[3pt]
$EM_2$ ($10^{53}$ cm$^{-3}$)& 
 $18.7_{-10.2}^{+10.7}$& $41.8_{-13.0}^{+13.8}$& $39.9_{-23.8}^{+11.7}$ & &
 $14.5_{-2.4}^{+2.4}$  & $16.2_{-4.6}^{+2.5}$  & $15.9_{-1.3}^{+1.3}$ \\[3pt]
O & 
 $0.76_{-0.33}^{+1.18}$& $0.33_{-0.13}^{+0.23}$& $0.34_{-0.10}^{+0.14}$ & &
 $1.56_{-0.84}^{+1.24}$& $0.13_{-0.13}^{+0.48}$& $0.75_{-0.26}^{+0.34}$\\[3pt]
Ne& 
 $0.99_{-0.38}^{+1.31}$& $0.48_{-0.16}^{+0.28}$& $0.51_{-0.11}^{+0.15}$ & &
 $2.11_{-1.15}^{+1.61}$& $0.76_{-0.40}^{+1.42}$& $2.71_{-0.64}^{+0.80}$ \\[3pt]
Mg& 
 $0.88_{-0.38}^{+1.24}$& $0.53_{-0.21}^{+0.35}$& $0.46_{-0.12}^{+0.17}$ & &
 $0.21_{-0.21}^{+0.93}$& $0.16_{-0.16}^{+0.54}$& $1.74_{-0.74}^{+0.90}$ \\[3pt]
Si& 
 $1.72_{-0.80}^{+2.37}$& $1.01_{-0.45}^{+0.73}$& $0.84_{-0.26}^{+0.32}$ & &
 $0.68_{-0.57}^{+0.76}$& $0.24_{-0.24}^{+0.53}$& $0.60_{-0.48}^{+0.55}$ \\[3pt]
S & 
 $= 1$                 & $= 1$                 & $= 1$                  & & 
 $0.10_{-0.10}^{+0.78}$& $0.35_{-0.35}^{+1.07}$& $0.00_{\dots}^{+0.50}$\\[3pt]
Fe& 
 $0.70_{-0.24}^{+0.87}$& $0.39_{-0.13}^{+0.21}$& $0.40_{-0.08}^{+0.10}$ & &
 $0.36_{-0.18}^{+0.23}$& $0.07_{-0.05}^{+0.24}$& $0.44_{-0.12}^{+0.13}$ \\[3pt]
$\chi_r^2$/d.o.f& 
 1.78/55               & 1.50/56               & 1.15/174 & & 
 0.54/97               & 0.87/104              & 0.67/288 \\[3pt]
$F_x^{\,a}$ ($10^{-12}$ erg~cm$^{-2}$~s$^{-1}$)& 
 4.35                  & 5.26                  & 5.12& & 
 1.44                  & 1.71                  & 1.59\\[3pt]
$L_x^{\,a}$ ($10^{31}$ erg~s$^{-1}$)& 
 6.45                  & 7.80                  & 7.59& & 
 2.14                  & 2.54                  & 2.36 \\
\noalign{\smallskip}\hline \noalign{\smallskip}
 &\multicolumn{3}{c}{\bf Source 3}& &\multicolumn{3}{c}{\bf Source 4} \\[6pt]
$T_1$(keV)& 
 $0.73_{-0.42}^{+0.10}$& $0.76_{-0.17}^{+0.15}$& $0.74_{-0.17}^{+0.08}$ & &
 $0.46_{-0.09}^{+0.33}$& $0.42_{-0.09}^{+0.25}$& $0.48_{-0.08}^{+0.18}$\\[3pt]
$T_2$(keV)& 
 $2.07_{-0.35}^{+0.16}$& $2.58_{-0.55}^{+1.46}$& $2.29_{-0.44}^{+0.28}$ & & 
 $2.86_{-0.63}^{+0.84}$& $2.29_{-0.37}^{+0.40}$& $2.55_{-0.23}^{+0.24}$ \\[3pt]
$EM_1$($10^{53}$cm$^{-3}$)& 
 $3.68_{-1.44}^{+2.57}$& $7.00_{-3.74}^{+5.44}$  & $4.54_{-1.09}^{+3.81}$ & & 
 $1.71_{-1.05}^{+8.20}$& $3.63_{-1.27}^{+3.98}$& $1.51_{-0.43}^{+0.80}$ \\[3pt]
$EM_2$($10^{53}$cm$^{-3}$)& 
 $9.21_{-1.36}^{+3.47}$& $6.61_{-1.41}^{+2.30}$& $7.49_{-1.56}^{+1.57}$ & & 
 $5.65_{-1.71}^{+1.81}$& $8.72_{-1.53}^{+1.93}$& $6.79_{-0.87}^{+0.98}$ \\[3pt]
O& 
 $0.58_{-0.51}^{+0.76}$& $0.49_{-0.39}^{+0.84}$& $0.54_{-0.36}^{+0.17}$ & & 
 $0.70_{-0.57}^{+1.09}$& $0.14_{-0.14}^{+0.39}$& $0.83_{-0.39}^{+0.73}$ \\[3pt]
Ne& 
 $0.84_{-0.64}^{+0.81}$& $0.96_{-0.53}^{+0.99}$& $1.26_{-0.55}^{+0.67}$ & & 
 $2.41_{-1.50}^{+2.35}$& $1.10_{-0.58}^{+1.02}$& $2.55_{-0.85}^{+1.08}$ \\[3pt]
Mg& 
 $= 0.20$              & $0.16_{-0.16}^{+0.50}$& $0.23_{-0.23}^{+0.29}$ & & 
 $1.27_{-1.01}^{+1.86}$& $0.40_{-0.40}^{+0.72}$& $0.68_{-0.61}^{+0.79}$ \\[3pt]
Si& 
 $0.09_{-0.09}^{+0.31}$& $0.35_{-0.29}^{+0.49}$& $0.06_{-0.06}^{+0.21}$ & & 
 $1.56_{-1.08}^{+1.54}$& $0.53_{-0.50}^{+0.57}$& $0.14_{-0.14}^{+0.47}$ \\[3pt]
S& 
 $0.30_{-0.30}^{+0.36}$& $0.87_{-0.67}^{+0.89}$& $0.18_{-0.18}^{+0.37}$ & & 
 $0.80_{-0.80}^{+1.50}$& $0.00_{\dots}^{+0.31}$& $0.00_{\dots}^{+0.42}$\\[3pt]
Fe& 
 $0.17_{-0.11}^{+0.10}$& $0.09_{-0.05}^{+0.14}$& $0.15_{-0.08}^{+0.10}$ & & 
 $0.28_{-0.20}^{+0.33}$& $0.10_{-0.07}^{+0.14}$& $0.34_{-0.14}^{+0.16}$ \\[3pt]
$\chi_r^2$/d.o.f& 
 0.69/69               & 0.55/57               & 0.64/202 & & 
 0.61/55               & 0.73/61               & 0.62/191\\[3pt]
$F_x^{\,a}$ ($10^{-13}$ erg~cm$^{-2}$~s$^{-1}$)&  
 9.64                  & 9.60                  & 8.83 & & 
 6.20                  & 7.86                  & 7.06 \\[3pt]
$L_x^{\,a}$ ($10^{31}$ erg~s$^{-1}$)& 
 1.43                  & 1.42                  & 1.31 & & 
 0.92                  & 1.17                  & 1.05 \\
\noalign{\smallskip}\hline\noalign{\smallskip}
\multicolumn{8}{l}{$^a$: unabsorbed flux and luminosity in the {\it ROSAT}
($0.1-2.4$ keV) band. $L_x$ has been computed assuming} \\
\multicolumn{8}{l}{$d=352$ pc for all
sources. Note that the luminosity of $\sigma$~Ori E might be higher by a
factor of 3.3 if its}\\
\multicolumn{8}{l}{distance is 640 pc, see Sect.~\ref{sec:soriE_disc} }
\end{tabular}
\label{tab:fit2T}
\end{table*}


PN and MOS spectra of these four sources have been extracted from the same
regions as the light curves; the background spectrum was extracted from a
nearby circular region of radius $24\arcsec$ free from other X-ray sources
and on the same CCD chip. Response matrices were generated for each source
using the standard SAS tasks. Spectra have been rebinned in order to have at
least 30 counts per bin, and were fitted in XSPEC v.11.2.0 in the energy
range $0.5-8$ keV, using a two-temperature APEC v.1.3.0 model with variable
element abundances. Since the hydrogen column density is not constrained by
the fit, it was kept fixed to the value $N_{\rm H} = 2.7 \times 10^{20}$
cm$^{-2}$, derived from the measured reddening $E(B-V)=0.05$ for
$\sigma$~Ori \citep{lee68,brown84}. The spectra of the four sources together
with the best-fit models are shown in Fig.~\ref{fig:epicspec}, and the
best-fit parameters are given in Table~\ref{tab:fit2T}. Abundances are
relative to the solar abundances by \citet{anders89}. Note that the spectral
fit for $\sigma$~Ori~E refers to the whole (quiescent + flare) observation,
since we will use it to correct the RGS spectrum obtained from the entire
observation.

Fig.~\ref{fig:epicspec} clearly shows that $\sigma$~Ori~AB has a softer
spectrum than the other three sources, indicating that it is much cooler
than the others, as expected for an early-type star. The fit indicates
temperatures of $\sim 0.1$ and $\sim 0.3$ keV (i.e. $\log T[{\rm K}] \sim
6.1$ and $\sim 6.5$). The two K stars in the field show instead hard
spectra, with temperature components of $0.5-0.7$ keV ($\log T[{\rm K}] \sim
6.8-6.9$) and $\sim 2-3$ keV ($\log T({\rm K}) \sim 7.4-7.5$) that are
typical of young active late-type stars. More complex is the case of the
B2Vp star $\sigma$~Ori~E, which is an early-type star but which has also
been caught during a flare. Its integrated spectrum is quite hard, and
clearly shows the iron complex at 6.7~keV. There is also evidence for an
excess emission at $\sim 6.4$~keV, likely due to the fluorescence line of
\ion{Fe}{i}, which is a signature of cool material close to the X-ray
source. 


\section{Results}
\label{sec:results}

\begin{figure*}
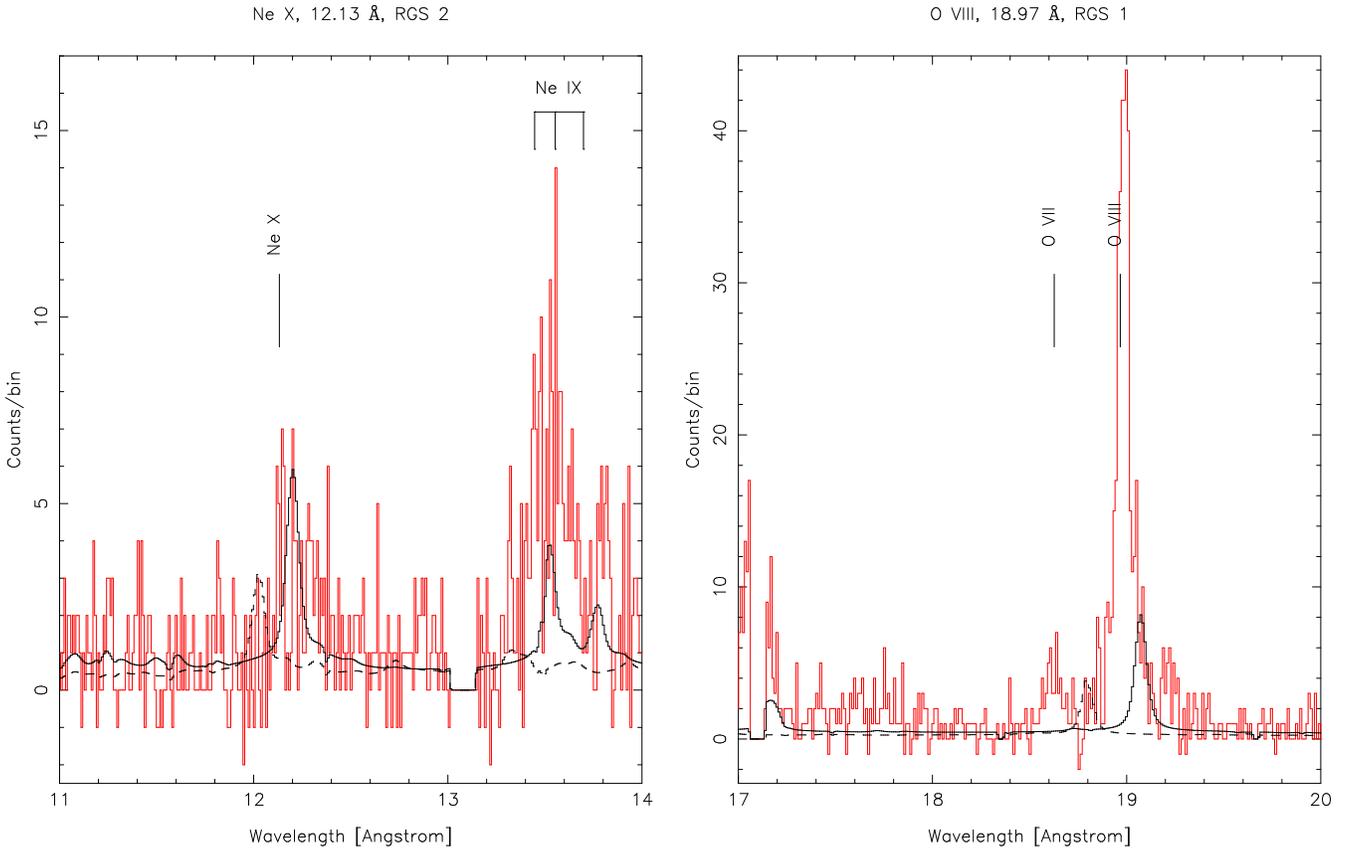

\resizebox{\hsize}{!}{\includegraphics[angle=270]{contamNeX.ps} 
\includegraphics[angle=270]{contamO8.ps}} 
\caption{Selected wavelength intervals of the RGS spectrum of
$\sigma$~Ori~AB indicating the contamination by other sources in the field.
Prominent lines of the main source are identified. The dark solid line
represents the RGS spectrum of $\sigma$~Ori~E modeled from its EPIC spectrum
and shifted according to its position with respect to the main source.
Similarly, the dashed line represents the expected RGS spectrum of source
\#3 (see Table~\ref{tab:sources}) appropriately shifted in wavelength as if
it were contaminating the spectrum of the main source (source \#4 has almost
the same shift and a similar spectrum). Note the unusual strength of the
intercombination line of \ion{Ne}{ix} at 13.699~\AA\ (see text).
\label{fig:contamination}}
\end{figure*}

\subsection{The RGS spectrum of $\sigma$~Ori~AB}
\label{sec:rgsAB}

As mentioned in the previous section, the RGS FOV contains other three
sources that might contribute to the RGS spectrum of $\sigma$~Ori~AB. While
the contamination from sources \#3 and \#4 is expected to be negligible, the
one from the flaring source $\sigma$~Ori~E might be quite significant. It is
therefore necessary to identify the emission of these sources in the
spectrum of the hot star to allow the study of its thermal structure and
wind properties. Since we have the EPIC spectra of all the bright sources in
the field, we can model their X-ray emission and predict their expected
contribution to the RGS spectra, once their emission is shifted according to
their angular position in the RGS FOV, as explained above. While the
cross-calibration between the EPIC and RGS detectors has still some
uncertainties \citep[up to $\sim 20$\%,][]{kirs03}, such approach can be
safely employed to identify the lines of the main source ($\sigma$~Ori~AB)
that are significantly contaminated by the emission of the other sources,
and to estimate the overall effect of the contamination. As shown by
Fig.~\ref{fig:epicspec} and Table~\ref{tab:fit2T}, all the secondary sources
are hotter than $\sigma$~Ori~AB, and have a lower count rate. As a
consequence, their contribution to the RGS spectrum is limited to the
continuum at short wavelengths and to two spectral lines, \ion{Ne}{x}
$\lambda$12.1321 and \ion{O}{viii} $\lambda$18.97, as shown in
Fig.~\ref{fig:contamination}. Moreover, as shown in
Fig.~\ref{fig:sigorifov}, the region employed for the extraction of the RGS
spectra does not include sources \#3 and \#4, and therefore their
contribution can be neglected, although, as a double check, we also tested
the expected position of their lines in the assumption that these two
sources were also contaminating the spectrum of the main source. In
particular, Fig.~\ref{fig:contamination} shows that the \ion{O}{viii} line
of the main source is contaminated only to a small extent by $\sigma$~Ori~E;
the \ion{Ne}{x} line, however, has a more complicated behaviour, with
$\sigma$~Ori~E providing $\sim 50$\% of the observed line flux in the
spectrum and preventing any possible measure of the \ion{Fe}{xxi} line at
$\sim 12.284$~\AA. In the case of sources \#3 and \#4, even if they were
included in the extraction region the contamination would be limited only to
the continuum, while no significant contribution is expected in any of the
lines (Fig.~\ref{fig:contamination}).

\begin{table*}
\caption{Measured RGS line fluxes$^a$ of $\sigma$~Ori~AB. $\log T_{\rm max}$
indicates the maximum temperature (K) of formation of the line (unweighted by
the EMD). ``Ratio'' is $\log (F_{\rm obs}/F_{\rm pred})$ for each line. Blends
amounting to more than 5\% of the total flux for each line are indicated}
\label{tab:fluxes}
\begin{tabular}{lrccrrl}
\hline \hline
 Ion & $\lambda_{\rm model}$ &  
 $\log T_{\mathrm {max}}$ & $F_{\mathrm {obs}}$ & S/N & ratio & Blends \\ 
\hline \noalign{\smallskip}
\ion{Ne}{x} & 12.1321 & 6.8 & 1.19e-14 &   3.9 & $-0.33$ & \ion{Ne}{x}
12.1321, 12.1375, \ion{Fe}{xvii} 12.1240 \\
No id. & 12.2800 &  & 1.91e-14 &   5.0 & \ldots &  \\
\ion{Ne}{ix} & 13.4473 & 6.6 & 2.71e-14 &   6.0 &  0.35 & \ion{Fe}{xviii} 13.3948, \ion{Fe}{xix} 13.4620 \\
No id. & 13.4970 &  & 8.81e-15 &   3.4 & \ldots &  \\
No id. & 13.5180 &  & 2.12e-14 &   5.4 & \ldots &  \\
\ion{Fe}{xvii} & 15.0140 & 6.7 & 8.81e-14 &  16.6 & $-0.17$ &  \\
\ion{O}{viii} & 15.1760 & 6.5 & 2.41e-14 &   8.4 &  0.00 & \ion{O}{viii} 15.1765 \\
\ion{Fe}{xvii} & 15.2610 & 6.7 & 4.39e-14 &  11.2 &  0.06 &  \\
\ion{O}{viii} & 16.0055 & 6.5 & 2.66e-14 &   9.2 & $-0.48$ & \ion{Fe}{xviii} 16.0040, \ion{O}{viii} 16.0067 \\
\ion{Fe}{xviii} & 16.0710 & 6.8 & 1.81e-14 &   7.7 &  0.18 &
\ion{Fe}{xviii} 16.0450, 16.1590, \ion{Fe}{xix} 16.1100 \\
\ion{Fe}{xvii} & 16.7800 & 6.7 & 4.21e-14 &  11.4 & $-0.16$ &  \\
\ion{Fe}{xvii} & 17.0510 & 6.7 & 1.68e-13 &  19.1 &  0.19 & \ion{Fe}{xvii} 17.0960 \\
\ion{O}{vii} & 18.6270 & 6.3 & 2.35e-14 &   8.7 &  0.02 &  \\
\ion{O}{viii} & 18.9670 & 6.4 & 2.13e-13 &  25.7 &  0.13 & \ion{O}{viii} 18.9725 \\
No id. & 19.2160 &  & 1.25e-14 &   4.5 & \ldots &  \\
\ion{O}{vii} & 21.6015 & 6.3 & 1.55e-13 &  15.4 & $-0.06$ &  \\
\ion{O}{vii} & 21.8036 & 6.3 & 1.61e-13 &  15.6 &  0.20 &  \\
No id. & 22.0977 &  & 2.55e-17 &   0.2 & \ldots &  \\
\ion{N}{vii} & 24.7792 & 6.3 & 3.11e-14 &   9.8 & $-0.01$ & \ion{N}{vii}  24.7846 \\
\ion{C}{vi} & 33.7342 & 6.1 & 7.13e-14 &  11.2 &  0.00 & \ion{C}{vi} 33.7396 \\
\hline \noalign{\smallskip}
\end{tabular}

{$^a$ Line fluxes in erg cm$^{-2}$ s$^{-1}$}
\end{table*}

Once the features corresponding to $\sigma$~Ori~E that contaminate the RGS
spectrum of the central star have been identified, we can measure the RGS1
and RGS2 line fluxes using a continuum derived from a global fit to the RGS
spectra that accounts for the contributions from the two sources
($\sigma$~Ori~AB and $\sigma$~Ori~E) to first approximation. Measured line
fluxes were then corrected for the distance to the star (352 pc) and
interstellar absorption, in order to obtain the fluxes emitted from the source
(Table~\ref{tab:fluxes}). 

\begin{figure}
\resizebox{\hsize}{!}{\includegraphics{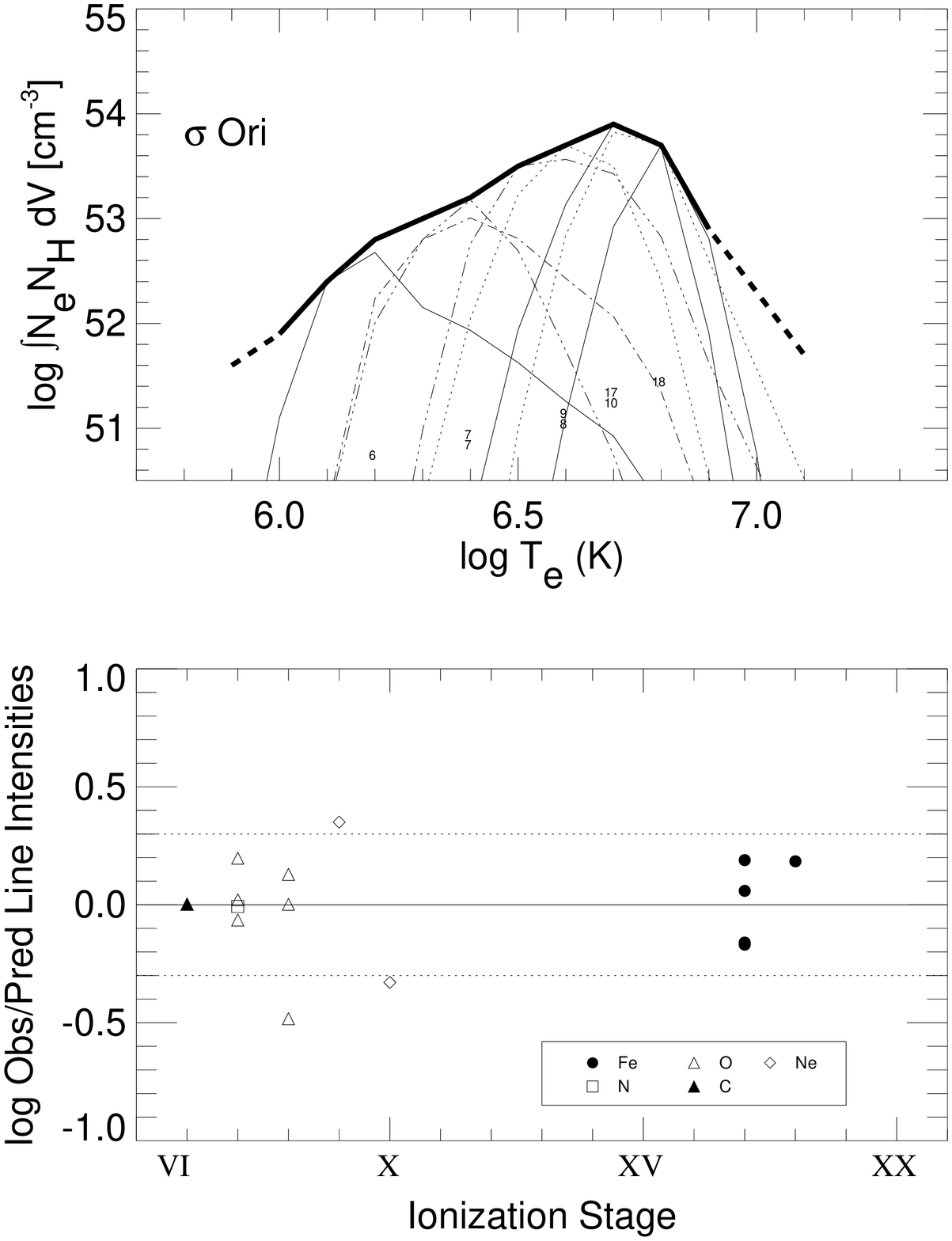}} 
\caption{Emission Measure Distribution (EMD) of $\sigma$~Ori~AB from the RGS
spectrum. {\em Upper panel}: EMD (thick line) and contributions of each
spectral line weighted by the EMD distribution (thin lines). {\em Lower
panel}: ratio of observed to predicted lines intensities for different
elements.\label{fig:sigoriemd}} 
\end{figure}

With the measured line fluxes we can determine the thermal structure of the
source by calculating the Emission Measure Distribution (EMD) as a function
of temperature, defined as $EM(T) = \int_{\Delta T} N_H N_e dV$ [cm$^{-3}$].
A line-based method has been employed in order to reconstruct the EMD of the
source and derive elemental abundances, following the procedure described
in \citet{sanz03}. The observed line fluxes, measured in RGS, are compared
to the predicted values for a given EMD; discrepancies found in the
comparison are reduced by correcting the adopted EMD and the values of the
abundances until a satisfactory result is found (i.e. when observed and
predicted line fluxes best agree). Since the lines emitted by different
elements have contributions that overlap in temperature only partially, some
uncertainty exists in the determination of the EMD and abundances in a case
like that of $\sigma$~Ori~AB, where the number of measured spectral lines is
small. The derived EMD is given in Fig.~\ref{fig:sigoriemd}: it peaks at
$\log T ({\rm K})\sim 6.7$, close to the results of the global fits to the
EPIC spectra. 

The abundances derived from the EMD reconstrunction are (in the usual
notation) [C/Fe]\,$=0.53 \pm 0.09$, [N/Fe]\,$=0.40 \pm 0.10$,
[O/Fe]\,$=0.45\pm 0.14$, [Ne/Fe]\,$=-0.50 \pm 0.41$. A value of
[Fe/H]\,$=-0.4$ (obtained from the EPIC/PN fit of $\sigma$~Ori~AB), has been
adopted in the EMD reconstruction. In the case of C and N it is not possible
to obtain reliable results from the EPIC analysis because the lines of these
elements are formed below 0.5~keV, a region affected by severe calibration
problems. While the Ne abundance is lower than the values obtained with EPIC
(Table~\ref{tab:fit2T}), O seems to be more abundant in the RGS spectrum than
in the EPIC spectra. Cross-calibration problems could be responsible for this
disagreement, as well as the assumption, in the EPIC fit, that $\sigma$~Ori~AB
is emitting only at two temperatures. The abundances derived from the multi-T
model (likely closer to the real plasma temperature structure) used in the RGS
analysis should be more reliable than those obtained with EPIC, where the low
spectral resolution allows us to apply only a 2-T fit. The results obtained
with RGS point towards a clear overabundance of C, N and O with respect to Fe. 

We have not detected any substantial blueshifts or Doppler broadenings in
the RGS lines, with an upper limit of $\sim$\,800~km~s$^{-1}$. Another
interesting feature of the RGS spectra is the observation of the He-like
triplets. These triplets, formed by the recombination ($r$),
intercombination ($i$), and forbidden ($f$) lines, are usually employed in
cool stars to derive the electron density in a collisionally excited plasma,
since the $f/i$ ratio is a decreasing function of density within a given
range of density values \citep{gab69}. However, it is also possible to have
a low $f/i$ ratio if a strong UV field is present close to the source
emitting the He-like triplets: in hot stars, which are strong UV sources, a
low $f/i$ ratio is thus indicative of proximity to the stellar surface
rather than of high density \citep[e.g.][]{cas01}. In the case of
$\sigma$~Ori~AB we observe the triplets of \ion{Ne}{ix}
($\lambda_f=13.699$~\AA, $\lambda_i=13.5531$~\AA), formed at $\log T({\rm
K})\sim 6.6$ and \ion{O}{vii} ($\lambda_f=22.0977$~\AA,
$\lambda_i=21.8036$~\AA), formed at $\log T({\rm K})\sim 6.3$. In both cases
the forbidden line is very weak if not completely absent (see
Figs.~\ref{fig:rgsspec} and \ref{fig:contamination}). This indicates that
the emission originates close to the star, which contradicts the usual
assumption of X-ray emission from wind shocks at large distances from the
star. We cannot exclude, however, a high density of the emitting region as
alternative, since the $f/i$ ratio allows us to put only an uninteresting
lower limit to density ($n_e \geq 10^8$~cm$^{-3}$) when a strong UV
radiation field is present. Note that the observed $f/i$ ratio could also be
the result of high densities, either close to the star (where the UV radiation
field is high), or at larger distances from the star. Hence, we cannot
distinguish which of the two mechanisms (UV radiation field, or high
densities) is responsible for the observed $f/i$ ratio.

\begin{table*}
\caption{Time-resolved spectroscopy of $\sigma$~Ori~E. Errors are 90\%
confidence ranges for one interesting parameter}
\begin{tabular}{lccrrrlcccc}
\hline \hline\noalign{\smallskip}
Interval& $T_1$& $T_2$& $EM_1/10^{53}$& $EM_2/10^{53}$& $Z/Z_\odot$& 
  $\chi^2_r$/d.o.f.& $F_x^{\,a}/10^{-12}$& $L_x^{\,a}/10^{31}$& $t_c^{\,b}$&
  $\Delta t$ \\
        & (keV)& (keV)& (cm$^{-3}$)   & (cm$^{-3}$)   &            
        &                  & (erg cm$^{-2}$ s$^{-1}$)& (erg s$^{-1}$)& (h)& 
  (h) \\
\noalign{\smallskip} \hline \noalign{\smallskip}
quiescent& $0.31_{-0.06}^{+0.10}$& $1.06_{-0.22}^{+0.21}$& 
  $7.09_{-2.33}^{+4.12}$ & $6.21_{-1.82}^{+3.59}$ &
  $0.09_{-0.06}^{+0.10}$& 1.00/26 & 0.59& 0.88& 12.3& 3.6 \\
\noalign{\smallskip}
peak     & $0.80_{-0.07}^{+0.15}$& $3.52_{-0.37}^{+0.40}$& 
  $3.62_{-1.14}^{+1.99}$ & $34.88_{-2.38}^{+3.74}$&
  $0.73_{-0.27}^{+0.31}$& 0.68/101& 3.06& 4.54& 15.5& 1.7 \\
\noalign{\smallskip}
decay  & $0.75_{-0.10}^{+0.07}$& $3.25_{-0.30}^{+0.41}$& 
  $6.35_{-2.30}^{+6.08}$ & $32.25_{-2.33}^{+2.49}$& 
  $0.30_{-0.06}^{+0.06}$& 0.91/129& 2.64& 3.91& 17.5& 2.5 \\
\noalign{\smallskip}\hline\noalign{\smallskip}
\multicolumn{10}{l}{$^a$: unabsorbed flux and luminosity in the {\it ROSAT}
($0.1-2.4$ keV) band. $L_{\rm X}$ has been computed assuming $d=352$ pc} \\
\multicolumn{10}{l}{$^b$: central time of the intervals in hours from
JD\,2452357}\\
\end{tabular}
\label{tab:soriE_evol}
\end{table*}


\begin{figure}
\resizebox{\hsize}{!}{\includegraphics{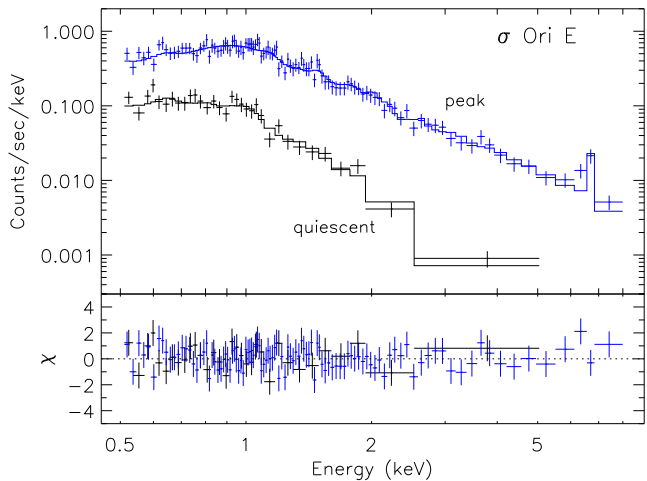}}
\caption{PN spectra of $\sigma$~Ori~E during quiescence and at the peak of the
flare. The best-fit model is also shown}
\label{fig:soriE_evol}
\end{figure}

\begin{figure}
\resizebox{\hsize}{!}{\includegraphics{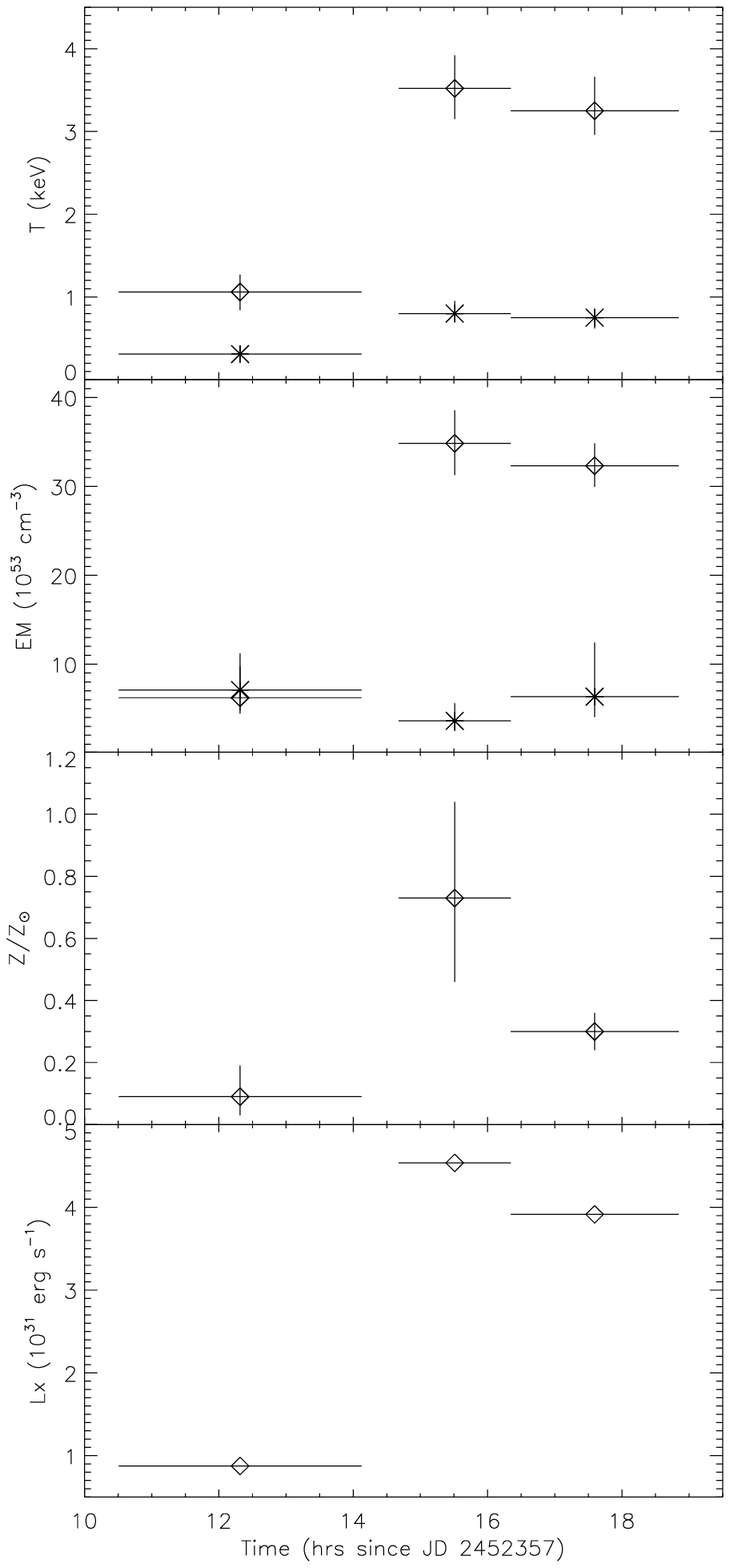}}
\caption{Time evolution of the best-fits parameters and of the X-ray
luminosity during the flare on $\sigma$~Ori~E}
\label{fig:soriE_ev_pars}
\end{figure}

\subsection{The flare on $\sigma$~Ori~E}
\label{sec:soriE_flare}

As mentioned in Sect.~\ref{sec:centr_src}, the most striking feature of our
observation is the flare observed on the hot star $\sigma$~Ori~E. The flare
lasted for $\ge 12$ hrs, as estimated by extrapolating the observed decay,
with a rise time of $\sim 1$~hr; the decay phase was observed only for the
first 6.7~hr. We have performed time-resolved spectral analysis of both the
quiescent emission and the flare at the peak and during the first part of
the decay; the chosen intervals are shown in Fig.~\ref{fig:lcurves}. Given
the low number of counts in each interval, we have only fitted the PN
spectra using a 2-temperature APEC model with variable global abundance. The
best-fit parameters are shown in Table~\ref{tab:soriE_evol}.
Fig.~\ref{fig:soriE_evol} shows the spectra obtained during quiescence and
at the peak of the flare, together with the corresponding best-fit models.
The estimated total energy released by the flare is $E \ga 10^{36}$ erg.

An interesting feature derived from this analysis is the preflare spectrum
of $\sigma$~Ori~E, which is quite unusual for a hot star and similar to
typical spectra of active cool stars (although not as hard as those of sources
\#3 and \#4, cf. Table~\ref{tab:fit2T}). In fact the source is much hotter
than $\sigma$~Ori~AB, with a comparable amount of material at temperatures of
0.3 and 1.1 keV ($\log T[{\rm K}]\sim 6.5$ and 7.1). During the flare, the
temperature of both components increases significantly (see
Fig.~\ref{fig:soriE_ev_pars}), with a large increase of the emission measure
of the hotter component ($EM_2/EM_1 \sim 10$). At the peak of the flare the
emitting plasma reaches a temperature of 3.5 keV ($\log T[{\rm K}] \sim 7.6$)
and there is a significant increase of the abundance, from $Z\sim 0.1 \,
Z_\odot$ during quiescence to $Z\sim 0.7 \, Z_\odot$ at the peak of the flare,
followed by a decrease during the decay. The temperatures and emission
measures do not decrease significantly during the first part of the decay,
suggesting the presence of a prolonged heating mechanism. This is consistent
with the observed behaviour of the flare light curve, which shows a bump $\sim
2$~hrs after the peak that might be due to a secondary flaring event.

The EPIC spectrum of $\sigma$~Ori~E (Figs.~\ref{fig:epicspec} and
\ref{fig:soriE_evol}) shows another interesting feature, i.e. the presence
of excess emission around $\sim 6.4$ keV, which can be attributed to the
\ion{Fe}{I} fluorescence line, indicating the presence of cooler
circumstellar material ionized by the flare. Another possible hint of the
presence of circumstellar material is given by the dips identified in the
light curve at $t\sim 16.5$ and 17.5 hrs (see Fig.~\ref{fig:lcurves}),
corresponding to a photometric phase $\phi=0.8$ \citep[computed according to
the ephemeris of][$T_0=$\,JD\,2442778.819, $P_{\rm ph}=1.19084$~d]{rei00}.
At this phase, \citet{rei00} have found maximum He absorption, which has
been attributed to the presence of absorbing circumstellar material along
the line of sight. It is therefore possible that these dips are due to
absorption of the X-ray emission by this material. However, we cannot
exclude that such dips might be simply due to intrinsic variability of the
flaring emission. The current uncertainties on the EPIC calibration at low
energies does not allow us to obtain accurate fits below 0.5~keV: we are
therefore unable to determine from the spectral fits a reliable value for
the column density, and therefore to detect possible changes in the amount
of absorbing material towards $\sigma$~Ori~E during the flare.


\section{Discussion}
\label{sec:disc}

The interpretation of the X-ray emission of hot stars has been changing in
the past few years as a consequence of new high-spectral resolution data
obtained with {\it Chandra} and XMM-{\it Newton}. It has been traditionally
believed that X-ray emission in these stars arises from shock heating in
their radiationally-driven massive winds which are unstable and form high
density blobs on which the high-velocity winds shock. The presence of winds
is clearly seen in the high-resolution spectra of the few cases studied so
far \citep[e.g.][]{wald01,cas01,kahn,sch03}, but the behavior of the He-like
triplets in some of these sources indicates that at least part of the
emission must originate from a distance too close to the star to be
compatible to wind shocks \citep[e.g.][]{wald01}. An alternative explanation
of the observed behavior of the He-like triplets, however, is the presence
of high electron densities that could result from magnetically confined
structures, giving similar $f/i$ ratios in the He-like triplets as for the
case in which the UV radiation field is high, i.e., close to the stellar
surface. Equally intriguiging for our understanding of X-ray emission from
hot stars is the detection reported here and by \citet[from ROSAT
observations]{gro04} of a flare from $\sigma$~Ori~E, which is a peculiar hot
star, with the reported presence of strong magnetic fields of $\sim 10$~kG
at photospheric level \citep{lan78}. In discussing the data that we have
collected with XMM-{\it Newton}, we will treat separately the case of
$\sigma$~Ori~AB, for which high resolution spectra are available, from that
of $\sigma$~Ori~E, which, besides the strong flare, also shows an unusually
hard (for a hot star) quiescent EPIC spectrum.

\subsection{$\sigma$~Ori~AB}
\label{sec:soriAB_disc}

The RGS spectrum of $\sigma$~Ori~AB shows several intriguing features. The
He-like triplets of O and Ne indicate either the presence of high densities
(implying magnetic confinement at small or large distances from the star) or
a strong UV radiation field (implying proximity of the emitting material to
the stellar surface where the winds, however, have not been accelerated
enough to form strong shocks). Moreover, the X-ray spectrum reveals no signs
of strong winds ($v\la 800$~km~s$^{-1}$), but the observed $\log L_{\rm
X}/L_{\rm bol} \sim -6.4$ has a typical value for hot stars
\citep[cf.][]{pal81}. It is not likely that a cool low-mass companion may
be responsible for this emission, since a young active cool star with such a
high $L_{\rm X}$ would also have a much hotter X-ray spectrum than observed 
(cf. the case of $\sigma$~Ori~E below).

The radiation pressure in a O9.5 dwarf star, such as $\sigma$~Ori~AB, is
expected to be lower than in the case of hot giants and supergiants, or of
hotter stars, but it should be strong enough to prevent the formation of
coronal loops close to the star. Therefore, it is necessary to have either a
very strong magnetic field that confines a small amount of coronal material
close to the star (where the UV radiation field is high enough to cause the
observed low value of the $f/i$ He-like ratio), or some magnetic confinement
causing high densities must occur at larger distances from the star, where
the UV radiation field is too weak to justify the observed He-like ratios.
Interactions with circumstellar material, or within the circumstellar
material itself, might play a role in this context, as suggested by the
analogy with what is observed in some cases for PMS low-mass stars and their
circumstellar disks. 

X-ray observations of cool young stars have shown in fact an interesting
difference between the Classical T Tauri Stars, CTTS, (that still possess a
disk of circumstellar material), and Weak-lined T Tauri Stars, WTTS, (with
no disk around the star). While WTTS have an X-ray emission that is typical
of active cool stars, with high densities ($\sim 10^{12.5}$~cm$^{-3}$) at
temperatures of $\sim 10$~MK, but lower densities ($\sim
10^{9.5}-10^{10.5}$~cm$^{-3}$) at $\sim 2$~MK \citep[e.g.,
AB~Dor,][]{sanz03}, the only case of a CTTS studied so far at high
resolution, TW Hya \citep{kas02}, shows high densities at $\sim 2$~MK ($\ga
10^{12.5}$~cm$^{-3}$), while keeping the similarity with WTTS for the rest
of its coronal emission. This different behaviour led to the conclusion that
interaction with the disk, possibly though accretion, may play an important
role in the X-ray emission of CTTS. An interesting question is whether a
similar mechanism might also be relevant for hot stars. At this stage, this
suggestion remains only speculative, but could resolve some of the
conflicting results emerging from the observations of $\sigma$~Ori~AB and
of other hot stars.

We have found in $\sigma$~Ori~AB higher abundances of C, N and O with
respect to Fe than in the Sun, which could be indicative of processed
material from the interior of the star. This is difficult to understand in a
young unevolved main-sequence star. A crucial question is whether these
anomalously high CNO abundances agree or not with the photospheric
abundances. No accurate measurements of the photospheric abundances of this
star have been reported to our knowledge.

Finally, as mentioned above, what we have called $\sigma$~Ori~AB in this
paper is actually a quite complex system, only partially resolved by our
XMM-{\it Newton} observations (and better resolved, but not completely, by
{\it Chandra}). The primary source is itself a close binary, with components
of similar spectral type. Other two components of the same system
($\sigma$~Ori C and D) are unresolved at our resolution: from our data
(Fig.~\ref{fig:sori_comp}), and from the {\it Chandra} observation, they
seem to contribute little to the total observed X-ray emission, but they
might contribute to the circumstellar material and, possibly, to the
magnetic field configuration of the whole region. Moreover, there is another
X-ray and IR source, the protoplanetary disk $\sigma$~Ori IRS1 mentioned
above, unresolved at our resolution, that further complicates the picture.
Observations of other hot stars at high resolution, including both giants
and main-sequence stars, are clearly required to clarify the mechanisms of
X-ray emission in early-type stars.

\subsection{$\sigma$~Ori~E}
\label{sec:soriE_disc}

The presence of a flare in $\sigma$~Ori~E challenges, if it originated in
the hot star itself, the theories that interpret the X-ray emission of
early-type stars as due to wind shocks, assuming that no stellar corona can
be present. Stellar flares are produced by magnetic reconnection where
magnetic energy is converted into plasma heating, radiative losses and mass
motions. Flares are common among active stars, and especially among Young
Stellar Objects (YSO) and T Tauri stars (TTS) that possess a high rotation
rate. These stars have typical X-ray luminosities that can be a factor $\sim
10^3 - 10^4$ higher than that of the Sun and produce flares that are orders
of magnitude stronger than solar flares. The possibility that the observed
flare on $\sigma$~Ori~E, and part of its quiescent emission, were due to an
optically unconspicous low-mass companion (of spectral type K or M) cannot
be excluded. On the other hand, $\sigma$~Ori~E is known to be a magnetic
star, with a global magnetic field of $\sim 10$ kG \citep{lan78}, so the
interpretation of this observation is far from being straightforward.

Early-type stars follow the relationship $\log L_{\rm X}/L_{\rm bol}\sim -7$
\citep{pal81}. However, in the case of our observation, the quiescent
emission of $\sigma$~Ori~E ($L_{\rm X}\sim 9\times 10^{30}$ erg~s$^{-1}$)
results in a $\log L_{\rm X}/L_{\rm bol}\sim -5.8$ \citep[using bolometric
corrections by][]{flo96}, more than an order of magnitude higher than
expected. If we consider that an active young star like AB~Dor (K2V) has an
average quiescent emission of $L_{\rm X}\sim 1.5\times 10^{30}$ erg~s$^{-1}$
\citep[cf.][]{sanz03}, part of the observed quiescent X-ray emission of
$\sigma$~Ori~E might indeed arise from a young late-type companion. On the
other hand, the peak flare luminosity ($L_{\rm X}\sim 4.5\times 10^{31}$
erg~s$^{-1}$) is consistent with that of large flares in young active stars.

A late-type companion (later than $\sim\,$K0) would be consistent with the
observed colors of $\sigma$~Ori~E ($V=6.54$, $V-K=-0.40$, $J-K=0.02$, after
correcting for interstellar absorption). A K0 star of age $\sim 3$~Myr has
in fact $V-K=2.03$, $J-K=0.55$ \citep{siess00} while a typical B2V star has
$V-K=-0.66$ and $J-K=-0.12$ \citep{cox00}. If $\sigma$~Ori~E has a K0
companion, having $V\sim 10.7$ at the cluster distance \citep{siess00}, the
inferred $K$ magnitudes for the B2V and the K0 star would be, respectively,
7.2 and 8.7, with a summed magnitude $K\sim 6.96$; similarly, in the J band
the summed magnitude would be $J\sim 6.94$. Therefore, the observed IR
colors are consistent with a companion of spectral type later than $\sim\,$K0.
Moreover, the expected bolometric luminosity of the companion would be such
that $\log L_{\rm X}/L_{\rm bol}\sim -3.4$, consistent with that of an active
star close to saturation.

\citet{gro04} recently reported the detection of a flare from $\sigma$~Ori~E
with ROSAT, and attributed it to $\sigma$~Ori~E itself, rather than to a
low-mass companion. The authors argue that the star does not belong to the
$\sigma$~Ori cluster, but it is a background object at a distance of 640~pc
\citep{hun89}, rather than the $\sim 350$~pc of the cluster. This would
increase the quiescent X-ray luminosity of the star, and the total energy of
the flare, by a factor of $\sim 3.3$, but it would not change the essential
point that any late-type companion of $\sigma$~Ori~E must be itself a very
young object, whether the star belongs or not to the cluster. They also
argue that no changes in the radial velocity of the primary star have been
detected with a velocity of less than $\sim 1$~km~s$^{-1}$ \citep{gro77}, as
expected from a low-mass companion. Finally, \citet{gro04} state that the UV
to IR fluxes of the star agree well with those expected from $\sigma$~Ori~E
alone. These arguments can be answered as follows: (i) the method employed
by \citet{hun89} to derive the stellar distance is based on a complex
analysis of the optical spectrum of the star, which can be affected by
several uncertainties related to the models used to interpret the spectra.
Moreover, as mentioned above, $\sigma$~Ori~E, because of its spectral type,
is certainly a young star, whether or not it belongs to the cluster, and the
same must be true for any late-type companion physically related to it (we
exclude the unlikely circumstance that the hot star and its hypothetical
late-type companion are the result of a chance alignement). (ii) The high
precision ($\sim 1$~km~s$^{-1}$) quoted by \citet{gro77} for the measurement
of radial velocity variations in $\sigma$~Ori~E is quite remarkable
considering its high projected rotational velocity ($v\sin i\sim
140$~km~s$^{-1}$) and the instrumentation typically employed in these
measurements. A more realistic value of at least 10~km~s$^{-1}$ accuracy
would be enough to miss the detection of a companion of $\sim 0.7 M_\odot$
orbiting around a star with $M\sim 9 M_\odot$ (expected for a B2V star).
(iii) The IR colors of $\sigma$~Ori~E are compatible with the presence of a
low-mass companion, as explained above. Besides, the presence of a low-mass
star in the optical spectrum of an early-type star would be virtually
undetectable, especially if the late-type star has a high rotational
velocity and some circumstellar material.

The X-ray spectra of $\sigma$~Ori~E obtained by us with XMM-{\it Newton}
provide additional arguments in favour of the presence of an unseen
late-type companion. The observed EPIC spectra (see
Fig.~\ref{fig:soriE_evol}) indicate, even during the pre-flare quiescent
phase, a quite hard emission, similar to that of cool active stars, such as
AB~Dor or the sources \#3 and \#4 in the same XMM/EPIC field, but quite
different from the soft spectrum of $\sigma$~Ori~AB (see
Fig.~\ref{fig:epicspec}). The metal abundance during quiescence is very low,
$Z\sim 0.1\,Z_\odot$, similar to what is commonly found in active stellar
coronae. During the flare a significant increase of the metal abundance, by a
factor of $\sim 7$, is observed at the peak, similarly to what is observed in
several flares on active late-type stars \citep[e.g.][]{fav99,gud01yy}.

The observation of the fluorescence \ion{Fe}{i} line emission in the flaring
spectrum suggests the presence of circumstellar material around the flaring
source. This circumstellar material would block part of the emission of a
late-type companion in the optical range, making its detection even more
difficult. 

Despite all these arguments, that lead us to postulate the presence of an
unseen active late-type companion, an origin of the flare from the hot star
itself cannot be excluded. Flares in hot stars have been reported for the
B2e star $\lambda$~Eri \citep{smith93} and the Herbig Be star MWC 297
\citep{ham00}, in addition to the flare reported by \citet{gro04} in
$\sigma$~Ori~E. It is not easy to explain the flaring emission with the
usual magnetic reconnection in coronal loops (which are not expected to be
present in early-type stars), but alternative scenarios of magnetic
reconnection can be considered that include interactions between the star
and circumstellar material and/or with stellar disks, as it has been
proposed for classical T Tauri stars and other YSOs \citep{kas02}. For the
flare reported here, there is a remarkable coincidence between the
photometric phase when maximum He absorption is observed \citep{rei00} and
the presence of two dips in the X-ray light curve of the flare that might be
interpreted as absorption features. If we assume that the flare originated
in $\sigma$~Ori~E itself, the flaring source should be situated between the
stellar surface and the circumstellar material in order to produce the
absorption features observed in the flare decay. This is in contradiction
with the hypothesis of a flare arising from the outer magnetosphere as has
been suggested by \citet{gro04}.


\section{Conclusions}\label{sec:conclusions}

We have analysed an XMM-{\it Newton} observation of the $\sigma$~Ori cluster
centered on the hot star $\sigma$~Ori~AB. Our results can be summarised as
follows:
\begin{enumerate}
\item we have detected 174 X-ray sources in the $\sigma$~Ori field (above a
significance threshold of $5\sigma$) of which 75 are identified as possible
cluster members;
\item we have detected 5 early-type members of the cluster as well as
many late-type stars down to the substellar limit;
\item we have discovered rotational modulation due to surface activity in a
K-type star of the cluster (source \#4);
\item we have detected both quiescent and flaring emission from the B2Vp
star $\sigma$~Ori~E and we have found that the quiescent emission has an
EPIC spectrum unusually hard for a hot star;
\item we have argued that the X-ray flare on $\sigma$~Ori~E, as well as most
of the quiescent emission from the star, originated from an unseen young
late-type companion, rather than from the hot star itself;
\item we have obtained a high-resolution RGS spectrum of the central object
$\sigma$~Ori~AB and we have been able to separate the emission of the
central star from the contributions of nearby sources;
\item we have found that the emission of $\sigma$~Ori~AB is steady (at a level
$\la 1$\%) and much softer than the one associated with the other sources,
consistently with a wind origin;
\item we have found, however, no evidence (with an upper limit of $\sim
800$~km~s$^{-1}$) for line broadenings and shifts as could be produced by
strong winds;
\item we have found a very low value of the \ion{O}{vii} forbidden to
intercombination line ratio in the hot star which is at variance with
current models of shock heating in stellar winds far from the star;
\item finally, we have found higher abundances of CNO elements with respect
to Fe in the wind of $\sigma$~Ori~AB that are difficult to understand in a
young unevolved star.
\end{enumerate}
A complete census of all X-ray sources detected in this observation and
belonging to the cluster will appear in a companion paper currently in
preparation.


\begin{acknowledgements}
JSF acknowledges support by the Marie Curie Fellowships Contract No.
HPMD-CT-2000-00013. EF and RP acknowledge partial support from the Italian
Space Agency (ASI) for data analysis. This research has made use of NASA's
Astrophysics Data System Abstract Service.
\end{acknowledgements}



\end{document}